\documentclass[preprintnumbers, twocolumn, showpacs]{revtex4}
\usepackage{amssymb,amsmath,graphics,graphicx} 

\begin{document}
\title{Intermingled basins in coupled Lorenz systems}

\author{Sabrina Camargo $^1$, Ricardo L. Viana $^2$, and Celia Anteneodo $^{1,3}$}
\affiliation{$^1$ Department of Physics, PUC-Rio, Rio de Janeiro, Brazil \\
$^2$ Department of Physics, Federal University of Paran\'a, Curitiba, Brazil \\
$^3$ National Institute of Science and Technology for Complex Systems, Rio de Janeiro, Brazil.}

\pacs{0.45.Xt,05.45.Df,05.45.Pq,05.45.-a}


\begin{abstract}
We consider a system of two identical linearly coupled Lorenz oscillators,
presenting synchronization of chaotic motion for a specified range of the
coupling strength. We verify the existence of global synchronization and 
 antisynchronization  attractors with intermingled basins of attraction, 
such that the basin of one attractor is riddled with holes belonging to the basin of the
other attractor and {\it vice versa}. We investigated this phenomenon by
verifying the fulfillment of the mathematical requirements for intermingled
basins, and also obtained scaling laws that characterize quantitatively the
riddling of both basins in this system.  
\end{abstract}

\date{\today}

\maketitle

\section{Introduction}

The Lorenz system  
\begin{equation}
\label{lorenzx} {\dot x} =
\alpha(y-x), \;\;\; {\dot y} = \beta x - y - xz,  \;\;\; {\dot z} = - \gamma z
+ xy \,,
\end{equation} 
\noindent
for $\alpha = 10$, $\beta = 28$, and $\gamma = 8/3$, displays  a chaotic attractor with 
the familiar butterfly-like shape \cite{lorenz} . 
It is often quoted as a paradigmatic system in nonlinear
dynamics, since it displays many interesting dynamical properties of chaotic
dissipative systems.  
Moreover its equations mimetize the dynamical behavior
expected to occur in some physically relevant systems, as convection rolls in
the atmosphere \cite{lorenz}, single-mode lasers \cite{haken}, and
segmented disk dynamos \cite{dynamo}. 
Coupled Lorenz systems could
arise as well in the mathematical modeling of related physical problems. 
The simplest case in the latter
category is the coupling of two identical Lorenz systems.

In identical coupled systems, even if chaotic, synchronization 
of trajectories may occur \cite{pecora}. 
This phenomenon has been studied for more than two decades, motivating  a
wealth of analytical, numerical, and even experimental results \cite{kurthsbook}. 
Synchronization of chaos, besides its own interest as a
mathematical problem, finds applications for instance in secure communications \cite{securecomm}. 
The chaotic nature of the dynamics of one of the systems can be exploited to code messages
which could be sent to an identical system through some form of coupling. If
the latter system is synchronized with the former, the message can be securely
uncoded.

For two completely synchronized systems, 
 either periodic or   chaotic,  their  dynamical variables are equal for all times. 
 On the other hand,  if instead of the difference, it is the sum
of some of their dynamical variables that vanishes, 
the two systems are said to antisynchronize.  
Due to phase-space symmetries, coupled Lorenz systems  can exhibit 
both synchronized and antisynchronized states. 
Then, for secure communications purposes, the existence of another, antisynchronized, state is 
in principle a
source of troubles since, depending on the initial condition, the receiver
system could be tuned to the antisynchronized attractor.  This situation can still be 
dramatically worsen when the riddling phenomenon occurs.

As a matter of fact, multistable dynamical systems typically 
have a very complicated structure of basins of attraction,  
that may be delimited by fractal  boundaries  \cite{viana}.  
Suppose, for instance, that a dynamical system has two attractors, with the corresponding basins of
attraction sharing a common basin boundary in the phase space. If a ball
centered at   a given  initial condition and with 
a radius equal to the uncertainty
level intercepts the basin boundary, we cannot say {\it a priori} which
attractor the system will asymptote to \cite{grebogi85}.
If that boundary is a curve, even if fractal,  
the final-state sensitivity
problem can be circumvented by decreasing the radius of the uncertainty ball
(this can be done in  experimental  or numerical settings, 
by increasing the precision in
determining the initial condition in phase space). 
However,  such reduction of uncertain initial 
conditions is not possible in the limit case in which the fractal boundary 
is area-filling, i.e.,     the (box-counting) dimension of
the basin boundary gets close to the dimension of the phase space itself \cite{alexander}.  
In that  limit case, the fraction of uncertain initial conditions 
will likely not decrease no matter how much we decrease the uncertainty  
balls of each initial condition.  
The latter situation occurs for riddled basins \cite{alexander}.

From the mathematical point of view, riddled basins are observed  
in dynamical systems that exhibit an invariant  smooth  hypersurface with a
chaotic attractor lying on it,   another asymptotic final state, out of the
invariant subspace, and  negative Lyapunov exponent transverse to 
the invariant subspace with positive finite-time fluctuations \cite{ott1,scaling,on-off}.   
Under the conditions above, riddling originates from the loss of transversal 
stability of  unstable periodic orbits embedded in the chaotic attractor \cite{bifurcation}, 
despite the attractor being transversely stable in average. 
In this context, attractors must be understood in the weak sense of Milnor \cite{milnor}.
The transition associated to the first unstable orbit on the attractor 
that losses transversal stability determines the riddling bifurcation (see for 
instance Ref. \cite{yanchuk03} for an overview). 
Depending on the way these orbits loss stability and even on the dynamics outside the 
invariant manifold, different bifurcation scenarios and different forms of riddled basins 
can occur  \cite{bifurcation,periodic1,yanchuk01,maistrenko} (to cite a few examples).

If riddled basins exist in a multistable chaotic system,
their final states are utterly  unpredictable, i.e. we cannot say - with any
degree of certainty - which attractor the system will evolve to for 
long times \cite{sommerer}. The situation, in this case, is similar to that for a random
process, for which there can only be determined a probability for predicting the
final state of the system. In fact, some  phenomena  
formerly attributed to random variations in initial conditions can be also
interpreted as a consequence of riddling \cite{tribolium}. 

The simplest case of riddling is when only one of the coexisting attractors 
have a riddled basin. 
However, when there is more than one invariant subspace, then more than one attractor can be riddled. 
In this case, the basin structure is called {\it intermingled} \cite{alexander}. 

The aim of the present work is precisely to show the existence of intermingled basins of
attraction for the synchronized and antisynchronized states of two coupled
Lorenz oscillators. In previous literature there are already clues of such phenomenon.
Kim and coworkers  \cite{kim}, in a work about anti-synchronization of coupled chaotic oscillators,   
 point  to the possibility of a riddled basin of synchronization in  coupled Lorenz systems, 
but without going further on that issue. 
Furthermore, a one-dimensional  reduction of the Lorenz system
(to a piecewise approximation to the well-known Lorenz map) was
low-dimensional enough for an analytical treatment to be feasible and   
show the riddling of the synchronization basin \cite{verges}. 
The verification of the transversal stability conditions 
through direct methods (i.e., by making a linear stability analysis of each
invariant subspace) is quite difficult in two coupled Lorenz systems, since the phase space
is six-dimensional. Then, we investigate those properties numerically. 
We also characterize quantitatively the riddled basins  by means of  the scaling laws giving the
probability of making wrong predictions on the final state of the system, with
respect to two quantities of interest: (i) the phase-space distance to the
invariant subspace; and (ii) the uncertainty radius for each initial condition \cite{ott2}. 
We have verified that, for both quantities, the probability
scales as a power-law, as required for riddled basins.

The rest of the paper is organized as follows: Section \ref{sec:coupledlorenz} describes the
coupled system of Lorenz oscillators, as well as the existence of both
synchronized and antisynchronized states. Section \ref{sec:basins} presents a preliminary
discussion of the basins of attraction of both the synchronized and
antisynchronized states. The mathematical properties required for riddled
basins and the necessary tools  are the object of Section \ref{sec:riddling}.
Section \ref{sec:scaling} discusses the  quantitative characterization of riddled basins
through scaling laws and the theoretical results supporting them.  
 The last Section contains our conclusions and final remarks.

\section{Coupled Lorenz systems}
\label{sec:coupledlorenz}
   
Many different coupling schemes are possible for two identical Lorenz systems 
\cite{karnatak}. We have chosen, for symmetry reasons, a diffusive coupling
through the $z$-variable, as follows 
\begin{equation}
\begin{array}{ll}
\dot{x}_1= & \alpha(y_1-x_1),\\[1.0pt]
\dot{y}_1= & \beta x_1-y_1-x_1z_1,\\[1.0pt]
\dot{z}_1= & -\gamma z_1+x_1y_1+\varepsilon(z_2-z_1),\\[2.0pt]
\dot{x}_2= & \alpha(y_2-x_2),\\[1.0pt]
\dot{y}_2= & \beta x_2-y_2-x_2z_2,\\[1.0pt]
\dot{z}_2= & -\gamma z_2+x_2y_2+\varepsilon(z_1-z_2),\\[1.0pt]
\end{array}
\label{eq:coupled_lorenz}
\end{equation}
where we will use the same values for
$\alpha$, $\beta$, and $\gamma$, as in the uncoupled case, and $\varepsilon$
is the coupling strength.

On considering the dynamical behavior of the coupled system, it is convenient
to perform the  changes of variables  
\begin{equation}
\begin{array}{ccc}
x=\dfrac{(x_2-x_1)}{2},\;& y=\dfrac{(y_2-y_1)}{2},\;&z=\dfrac{(z_2-z_1)}{2},\\[2.5pt]
X=\dfrac{(x_2+x_1)}{2},\;& Y=\dfrac{(y_2+y_1)}{2},\;&Z=\dfrac{(z_2+z_1)}{2},\\
\end{array}
\label{eq:change}
\end{equation}
after which the coupled system (\ref{eq:coupled_lorenz}) becomes 
\begin{equation}
\begin{array}{rcl}
\dot{x} &=& \alpha(y-x),\\
\dot{y} &=& \beta x -y-(Xz+Zx),\\
\dot{z} &=& -(\gamma +2\varepsilon)z+Xy+Yx,\\ [3.0pt]   
\dot{X} &=& \alpha(Y-X),\\
\dot{Y} &=& \beta X - Y  -(XZ+xz),\\
\dot{Z} &=& -\gamma Z+XY+xy\,.\\
\end{array}
\label{eq:new_coupled}
\end{equation}
Whenever more convenient to the analysis, 
we will refer either to the new or the old variables. 

From inspecting Eqs.~(\ref{eq:new_coupled}) there follows that the
dynamics of the coupled system is invariant with respect to the transformation
$(x,y,z) \to (-x,-y,-z)$. Hence the conditions $x=y=z=0$ define an invariant
subspace ${\cal M}_s$ : one initial condition that belongs to this subspace
generates a trajectory in phase space that remains in ${\cal M}_s$ for any
time. This three-dimensional subspace defines the complete (or global)
synchronization manifold characterized by $x_1=x_2$, $y_1=y_2$, $z_1=z_2$. 

The dynamics in the invariant subspace ${\cal M}_s$, described by the
variables $(X,Y,Z)$, is governed by the equations of the uncoupled Lorenz
system, hence there is a chaotic attractor ${\cal A}_s$   (butterfly-like shape)  
lying in ${\cal M}_s$. 

Analogously, due to the symmetry $(X,Y,z) \to (-X,-Y,-z)$, the  states  
for which $X=Y=z=0$ define another invariant subspace ${\cal M}_a$ (anti-synchronization 
manifold), associated to the attractor ${\cal A}_a$, in which $(x,y,Z)$
follows the dynamics of the uncoupled system, i.e. ${\cal A}_a$ is a 
Lorenz chaotic attractor in ${\cal M}_a$.

There are also other symmetries already present in the uncoupled Lorenz system. 
Notice in Eqs.~(\ref{eq:coupled_lorenz}) that either $(x_1,y_1)\to
(-x_1,-y_1)$ or $(x_2,y_2)\to (-x_2,-y_2)$ lead, to  
four-dimensional invariant subspaces,   
while both symmetries together lead to a two-dimensional invariant subspace 
with a saddle point  at the origin. 
Finally, included in this two-dimensional subspace, the lines at $z_1=z_2$ and $z_1=-z_2$ 
also represent invariant subsets.
We did not find any other relevant attractor other than ${\cal A}_s$ and ${\cal A}_a$, 
which are attractors of the dynamics in the respective subspaces 
${\cal M}_s$ and ${\cal M}_a$, and can become attractors for the whole phase space 
depending on their transversal stability.

\section{Basins of attraction}
\label{sec:basins}

In dynamical systems with more than one attractor, the corresponding basins 
 may have fractal boundaries and even more complicated structures
like the Wada property \cite{wada}. Accordingly, in the coupled Lorenz
system (\ref{eq:coupled_lorenz}), the two coexisting attractors representing
synchronized and antisynchronized states are expected to have such complex
basin boundary structure.

\begin{figure}[b!]
 \includegraphics[width=0.45\textwidth,clip]{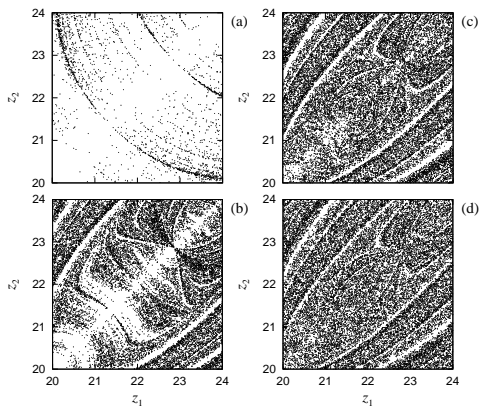}
\caption{\label{fig:basin}  Section at $x_1=x_2=y_1=y_2=1.0$ of the basins of  
 synchronization (white pixels) and  antisynchronization (black pixels) 
attractors of the coupled Lorenz system,  for  $\varepsilon =$ (a) $1.0$, 
(b) 2.0, (c) 2.5 and (d) 2.8. 
} 
\end{figure}

Since the phase space of the coupled system is six-dimensional, the
visualization of the basins of attraction depends on convenient phase space
sections  or projections. 
Figure \ref{fig:basin} shows a section  of the basin of the antisynchronization 
(synchronization) attractor  ${\cal A}_a$ (${\cal A}_s$), for different values of the 
coupling parameter.

Each initial condition was integrated using a fourth-order Runge-Kutta scheme
with fixed timestep $10^{-3}$ and for a time $t=10^3$, after which we determined
to what attractor the corresponding orbit has asymptoted \cite{exigence}. 
If an orbit has asymptoted to an antisynchronized (synchronized) state in ${\cal M}_a$ 
(${\cal M}_s$), its initial condition was painted black (white). Hence the area 
painted black (white) is a  numerical approximation of a section of the  basin  of  
attractor  ${\cal A}_a$ (${\cal A}_s$). 
We considered $10^5$ initial conditions
with $x_1=x_2=y_1=y_2=1.0$ while $z_1$ and $z_2$ were randomly chosen in the
interval $[20, 24)$ according to a uniform probability distribution.

\begin{figure}[h!]
\includegraphics[width=0.45\textwidth,clip]{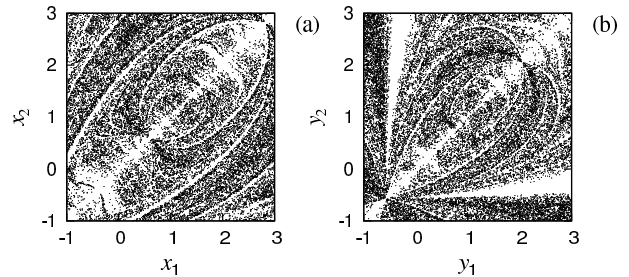}
\caption{\label{fig:basin2}  Section at $z_1=z_2=22.0$ and 
(a) $y_1=y_2=1.0$, $x_1,x_2$ random in $[-1,3)$, 
(b) $x_1=x_2=1.0$, $y_1,y_2$ random in $[-1,3)$ 
of the basins of  synchronization (white pixels) and  antisynchronization (black pixels) 
attractors of the coupled Lorenz system,  for  $\varepsilon = 2.0$. 
} 
\end{figure}

For instance, for a coupling strength  $\varepsilon =1.0$ [Fig. \ref{fig:basin}(a)], the section 
of the basin of  attractor ${\cal A}_a$ is a series
of thin filaments stemming from the diagonal. The filaments are
non-uniformly distributed and have a suggestive self-similar appearance. In
fact, successive magnifications of [Fig. \ref{fig:basin}(a)] reveal similar
patterns (see Ref. \cite{kim}).    
Such  scenario is also observed for other values of $\varepsilon$, as illustrated in 
Fig. \ref{fig:basin}(b-d), even if some features change with $\varepsilon$, such as the relative area of 
each basin, or the definition  of the tongues anchored in the diagonal.  
Let us note that  other cuts also display a tongue structure, as depicted in 
Fig.~\ref{fig:basin2} for $\varepsilon=2.0$.

\begin{figure}[h!]
\includegraphics[width=0.45\textwidth]{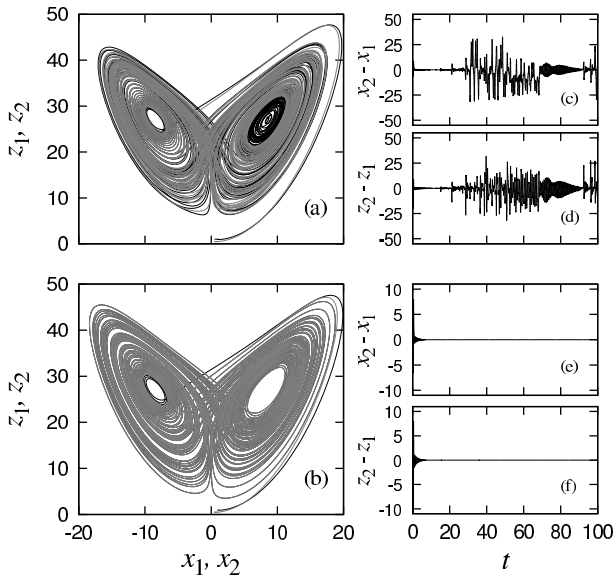}
\caption{\label{fig:butterfly} 
Trajectories of the coupled Lorenz system
for the same initial condition ($x_1=y_1=z_1=1.0$ and $x_2=y_2=z_2=0.5$) up to $t=100$ and 
different coupling values: (a)
$\varepsilon=0.5$; (b) $1.0$. In each case, the time evolution of the 
differences of coordinates are also shown (c)-(d) for $\varepsilon=0.5$ and (e)-(f) for 
$\varepsilon=1.0$.} 
\end{figure}

The structure of the basins of attraction is indeed expected to be altered by the
coupling strength. As an example, in Fig. \ref{fig:butterfly} we show that, 
for a given initial condition 
($x_1=y_1=z_1=1.0$ and $x_2=y_2=z_2=0.5$) integrated up to time $10^2$, the 
trajectories in the subspace of each oscillator are distinct for $\varepsilon=0.5$, 
while for $\varepsilon=1.0$  trajectories  tend to coincide due to synchronization.
In the latter case, the overlapping segments reproduce a cut of the familiar  
attractor of the single Lorenz system, since 
for synchronized orbits, the evolution proceeds towards the attractor in ${\cal M}_s$ 
which is defined by 
$x=y=z=0$, and $X=x_1=x_2$, $Y=y_1=y_2$, $Z=z_1=z_2$ 
 follow the dynamics of an uncoupled system, as described above.  
Differently, in the former case ($\varepsilon=0.5$), the trajectories of each system 
depart from those of the uncoupled system.  

Moreover, the observation of synchronized or antisynchronized states depends 
on the coupling strength. 
Recall that the  existence of ${\cal M}_s$
(i.e. the synchronized state being a possible solution of the coupled
equations) does not mean necessarily that synchronized states, 
and in particular states in its attractor ${\cal A}_s$,  can 
be observed in numerical simulations. This occurs only if there is 
transversal stability, in the sense that any infinitesimal displacement
along directions transversal to  ${\cal M}_s$ decays
exponentially with time. 

\begin{figure}[hb!]
\includegraphics[width=0.99\columnwidth]{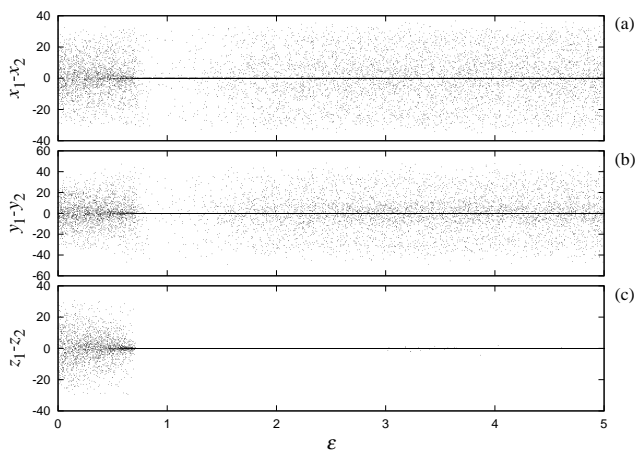}
\caption{\label{sync} Difference between the coordinates (a)
$x_1-x_2$; (b) $y_1-y_2$; (c) $z_1-z_2$  of two coupled Lorenz systems
 at time $t=10^3$, as a function of the coupling strength. 
One hundred  initial conditions were randomly chosen 
(as in Fig.~\ref{fig:basin}) for each value of $\varepsilon$ (varied in steps of  0.01).  
} 
\end{figure}

Let us remark that, due to the symmetry of the equations with respect to
synchronized/antisynchronized states, comments for   attractor ${\cal
A}_s$ are also valid for ${\cal A}_a$.

In order to visualize the existence of a transversely stable synchronization manifold, 
we consider the differences $x_1-x_2$, $y_1-y_2$, and $z_1-z_2$, which must
vanish if a synchronized attractor is achieved. For $\varepsilon \gtrsim 0.7$,  
 $z_1-z_2$  vanishes [Fig. \ref{sync}(c)], while the
other two differences may also vanish (global synchronization)
or not (local synchronization) [Fig. \ref{sync}(a) and (b)]. 
Similar plots are obtained for the sums $x_1+x_2$, $y_1+y_2$, 
indicating that  the basins of the synchronized and antisynchronized states are
complementary to each other.

\begin{figure}[ht!] 
\includegraphics[width=0.48\textwidth]{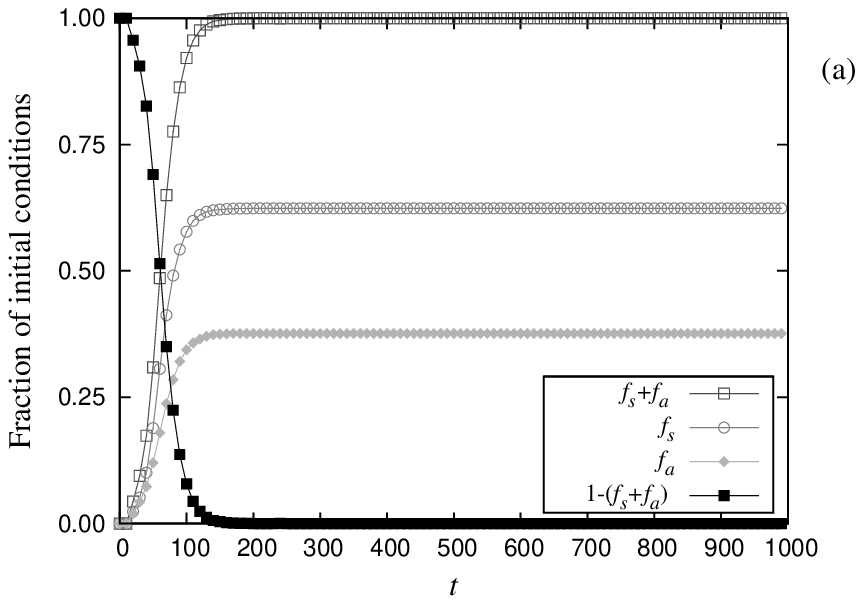}\\
\includegraphics[width=0.48\textwidth]{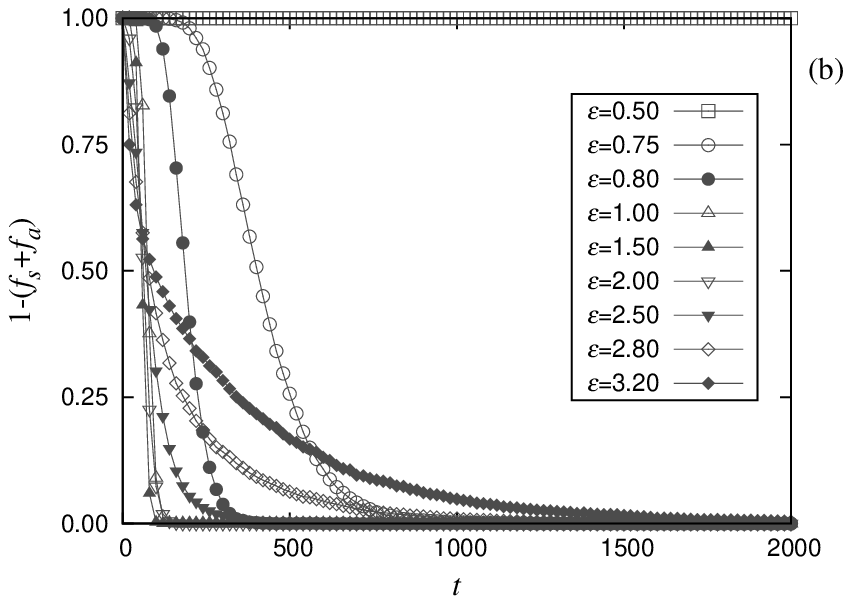}\\
\caption{\label{fracic} (a) Fraction of initial conditions (over a total of $10^4$)   
yielding trajectories asymptoting synchronized $f_s$ or antisynchronized
$f_{a}$ states as a function of time, for $\varepsilon = 2.0$. We
also indicated the fraction of trajectories reaching either state
($f_s+f_{a}$) or none of them ($1-f_s+f_{a}$). (b) Fraction of
initial conditions not reaching these states for different values of
the coupling strength.} 
\end{figure}

We did not find any relevant attractor for the coupled
system other than ${\cal A}_s$ and ${\cal A}_a$. 
Besides the symmetry considerations at the end of Sect. II, we  performed the 
following numerical experiment: we considered the
initial conditions used to plot the  sections   in Fig.~\ref{fig:basin} 
and, for each time $t$ we computed the fraction
of initial conditions that go either to ${\cal A}_s$ or ${\cal A}_a$ 
[Fig. \ref{fracic}(a)]. The sum of these fractions
rapidly  approaches  $100 \%$ 
[Fig. \ref{fracic}(a)], meaning that
the fraction of initial conditions that do not asymptote to them goes to zero
(filled squares in Fig. \ref{fracic}(a)), suggesting the existence of only two
attractors for the coupled system. This conclusion has been observed to hold
for $\varepsilon \gtrsim 0.7$ as illustrated in  [Fig. \ref{fracic}(b)]. 
Otherwise, neither  synchronized nor antisynchronized states are approached, 
as illustrated in Fig.~\ref{fig:butterfly}(c)-(d) for $\varepsilon=0.5$. 
(Therefore, basin diagrams as those shown in Fig. \ref{fig:basin} will be left blank).

\section{Riddled and intermingled basins}
\label{sec:riddling}

{The standard requirements for the existence of a riddled basin are the 
existence of  
(i) a smooth invariant subspace (of lower dimension than the phase-space) 
containing a chaotic attractor,   
(ii) another asymptotic final state (not necessarily chaotic) out of the
invariant subspace,  (iii) negativity of the  Lyapunov exponents transverse to 
the invariant subspace with (iv) positive finite-time fluctuations 
\cite{alexander,scaling,ott1}, which are associated to the transversal stability 
properties of unstable periodic orbits (UPOs) embedded in the attractor. 
For two symmetrically intermingled basins,  the requirements for mutual 
riddling can be summarized as follows:  

\begin{enumerate} 
\item There are invariant manifolds ${\cal S}_s$ and ${\cal S}_a$ contained in 
the phase space ${\cal H}$.  

\item The dynamics on each manifold ${\cal S}_{s,a}$ has a chaotic attractor.  

\item ${\cal S}_{s,a}$ are transversely stable, meaning that the largest 
transversal Lyapunov exponent $\lambda_\perp$ is negative.   

\item Although weak stability holds in average (condition 3),  
UPOs embedded in the chaotic attractor are transversely 
destabilized. 
\end{enumerate}

In Section \ref{sec:basins} we  showed that conditions 1 and 2 are fulfilled for the
coupled Lorenz systems. There exists two (three-dimensional) manifolds,  
${\cal S}_{s,a}={\cal M}_{s,a}$,    in the six-dimensional phase space. 
They are invariant since trajectories starting in each subspace will remain there forever. 
Because the dynamics in each subspace coincides with that of the uncoupled map, 
then, it will evolve towards  the respective well known Lorenz attractor 
 ${\cal A}_{s,a}$  lying in the corresponding manifold.

Moreover, for each invariant subspace, 
there are out of three transversal directions. 
Condition 3 means that the transverse Lyapunov exponents
of typical orbits lying in the invariant subspaces 
(${\cal M}_a$ and ${\cal M}_s$) are all negative. 
The point in parameter space where they become positive defines the blowout 
bifurcation \cite{ott1}. 
To investigate condition 3, it suffices to consider
the largest transversal exponent, denoted as $\lambda_\perp   =
\lim_{t\rightarrow\infty} {\tilde\lambda}_\perp({\bf x}_0,t) < 0$,  
where ${\bf x}_0$ is an initial condition on  the basin of attraction of either
${\cal A}_a$ or ${\cal A}_s$. 

We  computed  Lyapunov exponents using two different methods. 
 The Lyapunov spectrum  was obtained  
following the algorithm of Wolf {\it et al.} \cite{wolf.ea:85}, with a
Gram-Schmidt normalization step of $0.1$. We integrated Eqs.
(\ref{eq:coupled_lorenz}) using initial conditions given by
$x_1=y_1=x_2=y_2=1.0$ and $z_1,z_2$ were randomly chosen in the interval
$[20,24)$ from a uniform probability function, as in Fig. \ref{fig:basin}. 
These initial conditions lead to trajectories that asymptote 
to either ${\cal A}_s$ or ${\cal A}_a$.    As a matter of fact, this is not
relevant since both attractors have the same Lyapunov spectrum. 

The second method we used, and which can be applied to obtain only the largest 
transversal exponent, is to consider the time evolution of an infinitesimal
displacement along a direction transversal to the synchronized subspace ${\cal
M}_s$, which is given by \cite{viana}, 
\begin{equation}
\label{ftle_eq} 
\lambda_\perp = \lim_{t\rightarrow\infty} \frac{1}{t}
\ln \frac{\delta(t)}{\delta(0)}, 
\end{equation} 
\noindent 
where
$\delta(t)=\sqrt{(\delta x)^2+(\delta y)^2+(\delta z)^2}$ is the norm
of the transverse displacement, whose evolution is given by the
variational equations for $(x,y,z)$, setting $x=y=z=0$, i.e.,
\begin{equation}
\begin{array}{rcl}
\dot{\delta x} &=& \alpha(\delta y - \delta x),\\
\dot{\delta y} &=&  \beta\delta x -\delta y- X\delta z-Z\delta x,\\ 
\dot{\delta z} &=&  -(\gamma+2\varepsilon)\delta z+ X\delta y+Y \delta x,\\[2.5pt]
\dot{X} &=& \alpha(Y-X),\\ 
\dot{Y} &=& \beta X - Y  -XZ, \\ 
\dot{Z} &=& -\gamma  Z  + XY \,. 
\end{array}
\label{eq:transversal_eq}
\end{equation}

\begin{figure}[b] 
\includegraphics[width=0.94\columnwidth]{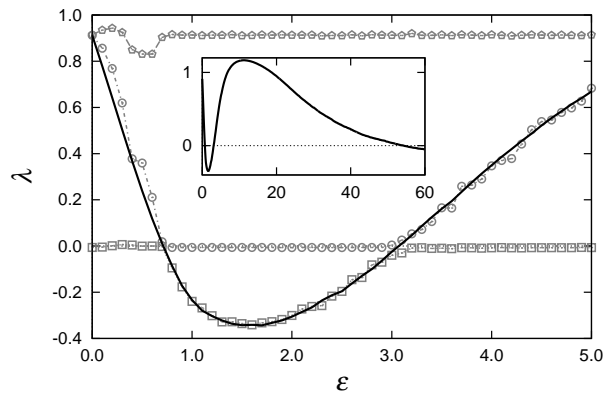}
\caption{\label{wolf_delta}  
The three largest Lyapunov
exponents of the coupled Lorenz system as a function of the coupling
strength. Gray symbols correspond to the algorithm by Wolf  {\it et al.}, 
whereas the  thick black curve is the result of the variational equations
(\ref{eq:transversal_eq}). Inset:   the largest exponent for a wider 
range of $\varepsilon$.} 
\end{figure}

Figure \ref{wolf_delta} shows (in gray symbols) the three largest
(infinite-time) Lyapunov exponents as a function of the coupling strength
$\varepsilon$, obtained by means of the algorithm by Wolf  {\it et al.}, 
and the largest transversal
exponent given by (\ref{ftle_eq}) is indicated by a thick black line. One of
the exponents is always zero, corresponding to displacements along the
trajectory. The largest exponent is practically always equal to $\sim 
0.9$ and corresponds to the chaotic dynamics on ${\cal A}_s$ 
(${\cal A}_a$). The third exponent is
the largest transversal exponent which we focus our attention on, both methods 
being in good accord in the region of interest 
(as shown in Fig. \ref{wolf_delta}).  
For the chaotic attractors in both synchronization and antisynchronization
manifolds, the largest transversal exponent vanishes, changing sign, at 
$\varepsilon_1 \approx 0.714\pm0.005$ and $\varepsilon_2 \approx 3.061\pm 0.005$, 
defining the critical points of the blowout bifurcation. 
(There is also another critical value for large $\varepsilon$, 
as can be seen in the inset of Fig. \ref{wolf_delta}, 
but we will restrict our analysis to the lower range only). 
The largest transversal exponent is negative for 
$\varepsilon_1 < \varepsilon < \varepsilon_2$, hence, in that   interval,   
condition 3 for intermingled basins is fulfilled.  
However, while the invariant subspaces ${\cal M}_{s,a}$ are stable in average, 
with negative transversal Lyapunov exponents, 
there may be particular unstable periodic orbits embedded in the chaotic attractors 
${\cal A}_{s,a}$ that are also transversely unstable,  
with positive largest transversal Lyapunov exponent \cite{heagy} (condition 4). 
When trajectories come close to these unstable orbits, they will be repelled from 
the vicinity of the attractor. 
This will be reflected in positive values of the  finite-time largest transversal Lyapunov 
exponent \cite{ott1,scaling,on-off}.  
Then we numerically computed  
the finite-time largest transversal Lyapunov exponents 
${\tilde\lambda}_\perp({\bf x}_0,t)$, 
by means of Eq.~(\ref{ftle_eq}) but at finite $t$. 
For large enough $t$,  one recovers the  infinite time   
exponent $\lambda_\perp$, which does not 
depend on ${\bf x}_0$, for   almost all initial conditions 
 in the attractors  ${\cal A}_{s,a}$, in contrast to the finite-time 
 ones that may depend on the initial condition.

We quantify the contributions of the finite-time largest transversal exponent
by obtaining a numerical approximation to the corresponding probability
distribution function $P({\tilde\lambda}_\perp({\bf x}_0,t))$. 
 We considered a large number of points in ${\cal M}_s$ 
(with $x=y=z=0$, $X=Y=1.0$, and $Z$ randomly chosen in the interval $[20,24)$), 
and  discarded a transient.
These were the initial conditions used to compute 
the time-$t$ largest transversal Lyapunov exponents.
%
Alternatively, we generated a single long chaotic trajectory (after the transient has elapsed)  
and divided it into time-$t$ segments, using then the ergodicity of 
the dynamics   to ensure that 
the conditions are randomly chosen according to the natural measure of the
attractor. The results were essentially the same.

\begin{figure}[h] 
\includegraphics[width=0.94\columnwidth]{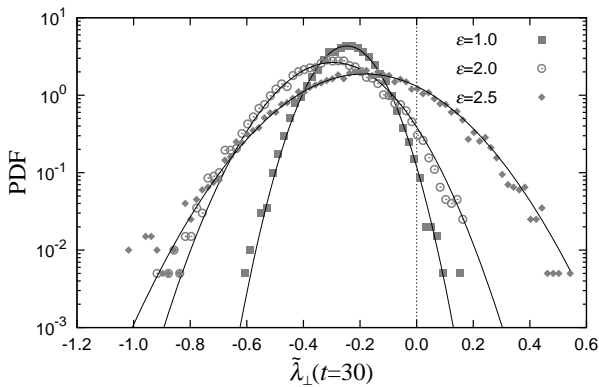}
\caption{\label{histo} Probability density functions of the time-$30$
largest transversal Lyapunov exponent for different values of
$\varepsilon$. Initial conditions were taken as in Fig. \ref{fig:basin}. 
The full lines are Gaussian fits.} 
\end{figure}

Figure \ref{histo} shows probability distribution functions 
(when $t=30$) for different values of the coupling strength. In all the 
considered cases the distribution is nearly Gaussian and presents positive 
tails. Then, we  computed the positive fraction of finite-time exponents,  
$\varphi(t) = \int_0^\infty P({\tilde\lambda}_\perp(t)) d{\tilde\lambda}_\perp(t) > 0$.  
The positive fraction is plotted in Fig. \ref{frac} as a function 
of the coupling strength for different values of the time-$t$
interval used to sample the finite-time exponents.

\begin{figure}[h] 
\includegraphics[width=0.94\columnwidth]{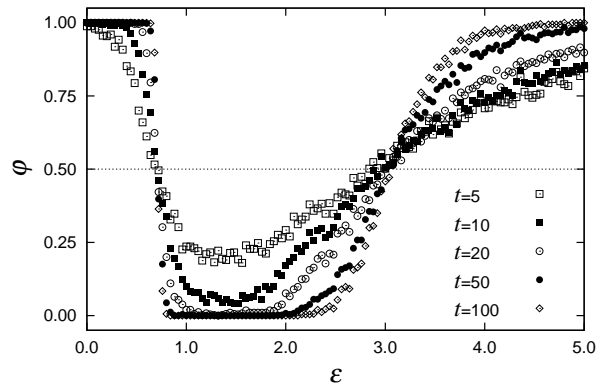}
\caption{\label{frac} Positive fraction of the largest time-$t$
transversal Lyapunov exponent as a function of the coupling strength,
for different values of $t$.} 
\end{figure}

For $\varepsilon \lesssim  \varepsilon_1$, finite-time exponents soon become  positive. 
This is in agreement with the positivity of
the infinite-time exponent shown in Fig. \ref{wolf_delta} for this region, and 
also with the fact that trajectories do not approach the invariant subspaces, but are soon 
repelled, as already noted when we tried to plot the basins in Section \ref{sec:basins}, 
which is not possible for that parameter range. 
The positive fraction drops rapidly to $50 \%$, which corresponds to the case
for which the infinite-time exponent vanishes (consistent with symmetric 
$P({\tilde\lambda}_\perp(t))$), and then drops below $50 \%$,
when the infinite-time exponent is negative. 
For $\varepsilon\simeq 1.4$, the positive fraction is minimal.  
For larger values of $\varepsilon$, it increases, crossing the  $50 \%$ level 
again for $\varepsilon \simeq \varepsilon_2$, 
where the infinite-time exponent is again zero at that point  (see Fig. \ref{wolf_delta}). 
After that, the positive fraction tends  to 1 gently with
$\varepsilon$, yielding a positive infinite-time exponent. 
This smooth behavior, different from the abrupt one in the lower limit of the 
region of negative $\lambda_\perp$, 
is consistent with the observation,  for $\varepsilon >\varepsilon_2$, 
of a basin structure reminiscent of those in Fig. \ref{fig:basin}, although the 
filaments from the diagonal are not neat. 
Even if trajectories are ultimately repelled, they can spend long time intervals close to 
each attractor.

The fraction of positive finite-time exponents is non-null. 
However, for the range  $\varepsilon_1<\varepsilon<\varepsilon_2$, 
that fraction decreases with $t$, as expected because the distribution of finite-time exponents 
collapses towards  a Dirac delta centered at $\lambda_\perp$ in the long time limit. 
The decay is exponential, 
the faster, the closer to the minimum at $\varepsilon\simeq 1.4$. 
Hence, the absence of an abrupt decay of the positive fraction indicates a 
nonvanishing fraction for finite times. 
Then, from this analysis, condition 4 cannot discarded for any range within the interval 
$\varepsilon_1 < \varepsilon <\varepsilon_2$. 
This suggests that at least one of UPOs  
should be transversely unstable in that interval. 

\begin{figure}[hb] 
\includegraphics*[bb=75 80 510 800, width=0.9\columnwidth]{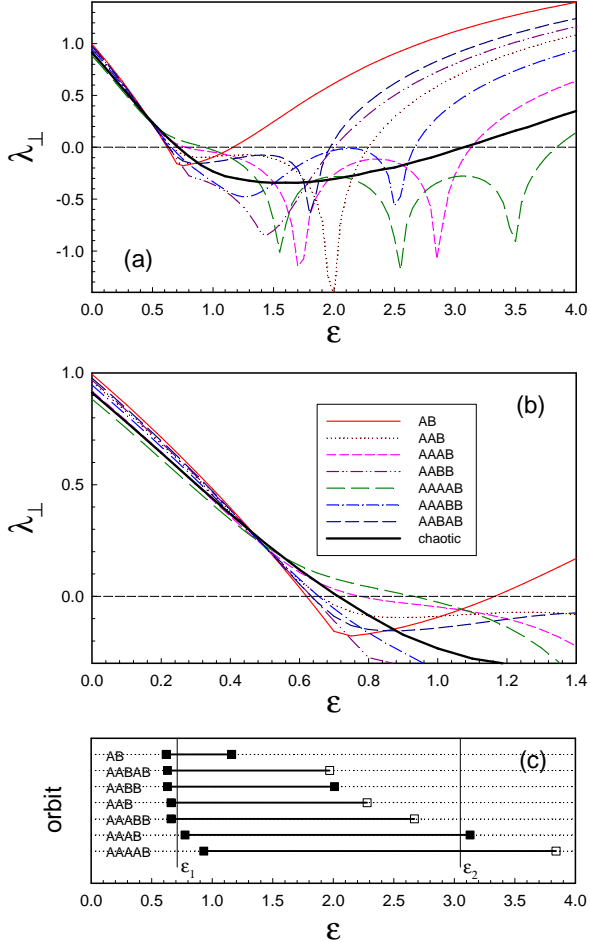}
\caption{ \label{fig:upos} (Color online)
(a) Largest transversal Lyapunov exponent, as a function of 
 $\varepsilon$, for the particular unstable periodic 
orbits (UPOs) embedded in the Lorenz chaotic attractor (up to period 5), 
indicated on the figure  by means of the sequence of symbols A, B 
denoting the turns around each unstable fixed points  
C$^+$ and C$^-$ of the Lorenz system. 
The curve for typical chaotic trajectories in the attractor is also shown.  
(b) Magnification of panel (a). 
(c) Stability intervals for each UPO, in order of increasing 
stability at $\varepsilon=0$ from top to bottom: stable (full segment), unstable (dotted).
The vertical lines indicate $\varepsilon_1$ and $\varepsilon_2$. 
The symbols delimiting the stability intervals correspond to $\mu=1$ (full) and $-1$ (hollow).
} 
\end{figure}

Then we  inspected the transversal stability of those orbits 
along the lines of  periodic orbits threshold theory \cite{heagy}. 
Once localization in phase space and periods of  low period UPOs 
are available in the literature for the Lorenz system \cite{upos},  
we computed Floquet multipliers \cite{heagy,trevisan}. 
Namely, we integrated Eqs.~(\ref{eq:transversal_eq}), to obtain the matrix 
$\mathbf{Q}$ such that $\vec{\delta}(\tau)=\mathbf{Q}\vec{\delta}(0)$, with $\tau$ 
the time period of the orbit and   
$\vec\delta(t)=(\delta x,\delta y,\delta z)$ the column vector of transverse deviations. 
The eigenvalue of $\mathbf{Q}$, $\mu$,  with maximal  modulus 
furnishes $\lambda_\perp=\ln|\mu|/\tau$, for a particular periodic orbit. 
Fig.~\ref{fig:upos} shows the behavior of  $\lambda_\perp$  as a function 
of $\varepsilon$ for particular UPOs, up to period 5. 
UPOs are labeled by means of the sequence of symbols A, B denoting the turns 
around each unstable fixed point C$^+$ and C$^-$ of the Lorenz system. 
Symmetric orbits obtained by exchanging A$\leftrightarrow$B or with cyclic symmetry    
were omitted.

One observes that the lowest period orbit AB (period 2) 
appears to be the first in destabilizing the  vicinity of $\varepsilon_2$, hence 
defining a riddling bifurcation.  
Then, between this point and $\varepsilon_2$ riddling can occur. 
This interval covers most of the range $\varepsilon_1 < \varepsilon <\varepsilon_2$, 
except for a very small interval in the vicinity of $\varepsilon_1$. 
However, note that orbits of the type A$^n$B, with $n=1,2,\ldots$, have a
maximal transversal Lyapunov 
exponents that increases with $n$  in the vicinity of  $\varepsilon_1$, hence 
shrinking the remaining small region of stability around $\varepsilon\simeq 1$.  
To confirm whether this  region of strong stability 
(with no transversely unstable orbits) actually disappears  
would require the analysis of higher period orbits, a hard task for this system, 
since the number of UPOs increases exponentially with the integer period.

Near the blowout bifurcation at $\varepsilon_1$, the
low-period UPOs (up to period 5) destabilize for coupling strength either  
weaker or stronger than the critical value, but close to it.  
Let us remark that, differently to the coupled R\"ossler system studied  
by Heagy et al. \cite{heagy}, here the ordering of the exponents of the 
lowest period orbits in the neighborhood of $\varepsilon_1$ is 
inverted with respect to the uncoupled case  as depicted  in Fig. \ref{fig:upos}. 
This implies that paradoxically the most stable orbits in the attractor are those 
responsible for the transversal destabilization in this parameter region. 
A similar inversion occurs on some domains of the parameter space of a system of 
symmetrically coupled  R\"ossler oscillators\cite{yanchuk03,yanchuk01}.
This characteristic turns difficult the determination of the riddling bifurcation 
(first destabilized orbit) related to the blowout at $\varepsilon_1$, 
apparently triggered by higher period orbits. 

Furthermore, our outcomes point to a different nature of the blowout bifurcations at 
$\varepsilon_1$ and $\varepsilon_2$.  
Nearby $\varepsilon_1$, UPOs destabilize in its vicinity. Moreover, for all the 
analyzed orbits, the multiplier $\mu$ crosses the circle $|\mu|=1$ 
along the real positive semi-axis (associated to a pitchfork bifurcation). 
This is in contrast to the scenario at $\varepsilon_2$, where there are 
orbits destabilizing far from $\varepsilon_2$ and with multiplier $\mu$ either 
+1 or $-1$. In particular, the first orbit AB loses stability with $\mu=1$.
The  differences are consistent with the picture given by   
finite-time exponents, for instance in connection with Fig. \ref{frac}, 
where a much abrupt behavior of the positive fraction was encountered near $\varepsilon_1$.

The intervals where riddling can occur are delimited on one side by the 
blowout bifurcation and on the other by the riddling bifurcation. 
In our case there are two of such intervals and they apparently overlap, 
such that  at least one UPO  has lost transversal stability 
 in the full range  $\varepsilon_1 < \varepsilon < \varepsilon_2$,  
although this would have to be confirmed by the analysis of high period orbits,  
it  is  supported by the analysis of finite-time Lyapunov exponents.

\section{Scaling laws for riddled basins}
\label{sec:scaling}

In this Section, we will focus on the determination of the scaling properties of 
the basins, which provide a measure of their structure\cite{scaling}.   
%
Let us focus on the black filaments in Fig. \ref{fig:basin}(b), which 
belong to the basin of the antisynchronization attractor. 
They are anchored at the  diagonal line, which is a cut of the 
synchronization manifold  ${\cal M}_s$, given by $x=y=z=0$  and   
 containing a chaotic attractor ${\cal A}_s$, 
while the antisynchronization attractor ${\cal A}_a$ lies elsewhere.
In Fig. \ref{sca1}(a) we portrait a schematic picture of
that structure. 
The filaments of the basin of ${\cal A}_a$ are tongues anchored at points of
${\cal A}_s$, and the complement of the filament set belongs to the basin of
${\cal A}_s$. If an initial condition starts within any of these narrow
tongues, even if it is very close to ${\cal A}_s$, the resulting trajectory
will asymptote to the other attractor.  

The set of basin filaments for ${\cal A}_a$ is expected to be self-similar by
quite general grounds.  
Once the riddling bifurcation occurs for a given periodic orbit, it also
occurs for every preimage of this orbit, yielding a dense set of tongue-like
sets anchored at the corresponding preimages on ${\cal A}_s$. The tongue-like
shape is a consequence of the nonlinear terms in the  equations describing
the transversal dynamics.  
The characteristic feature of riddling is that those
tongues have widths that tend to zero as we approach   ${\cal A}_s$. Hence
the basin of ${\cal A}_s$ always contains pieces of the basin of the other attractor,
regardless the transversal distance to ${\cal A}_s$, so forming a fine
structure of basin filaments (the same applying to ${\cal A}_a$). 

\begin{figure}[h] 
\includegraphics[width=0.75\columnwidth]{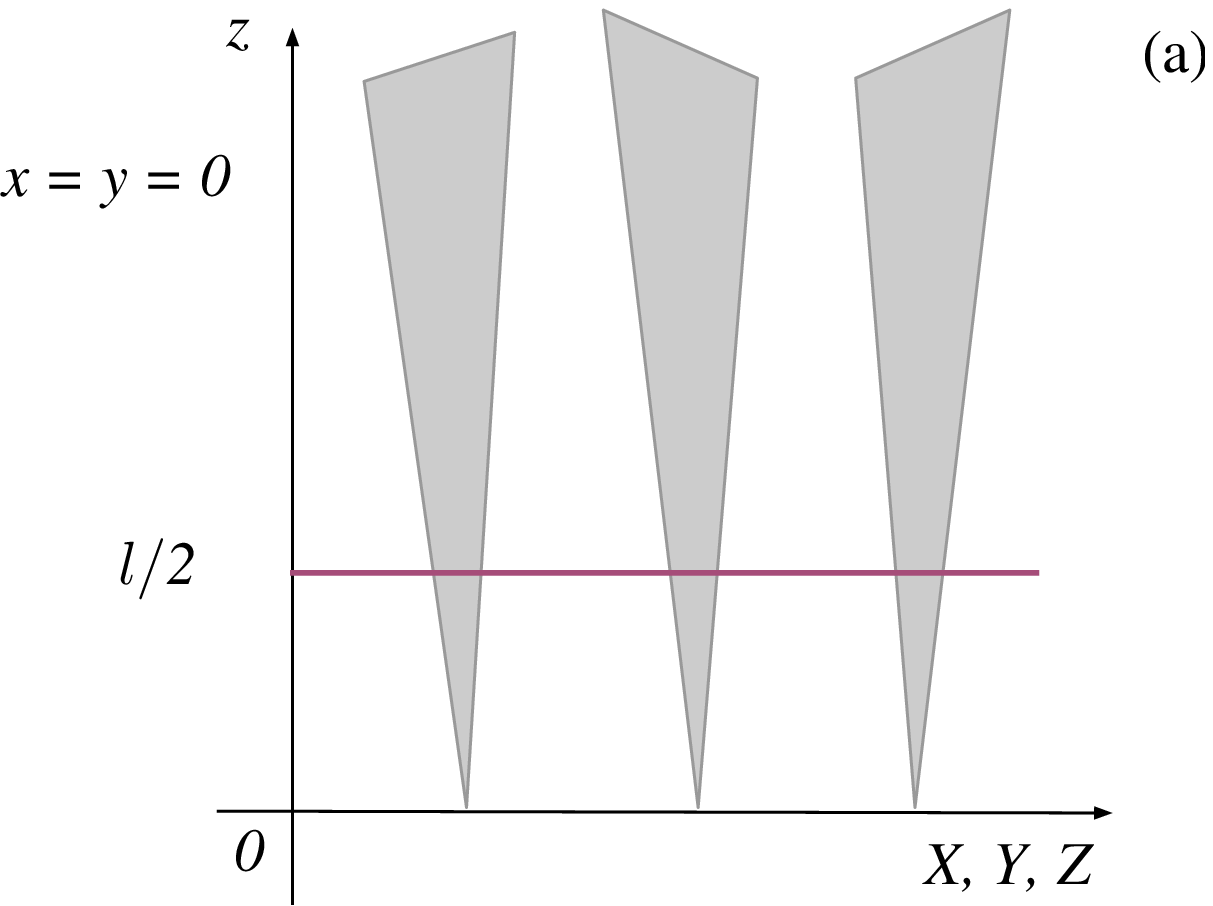}\\[5mm]
\includegraphics[width=0.8\columnwidth]{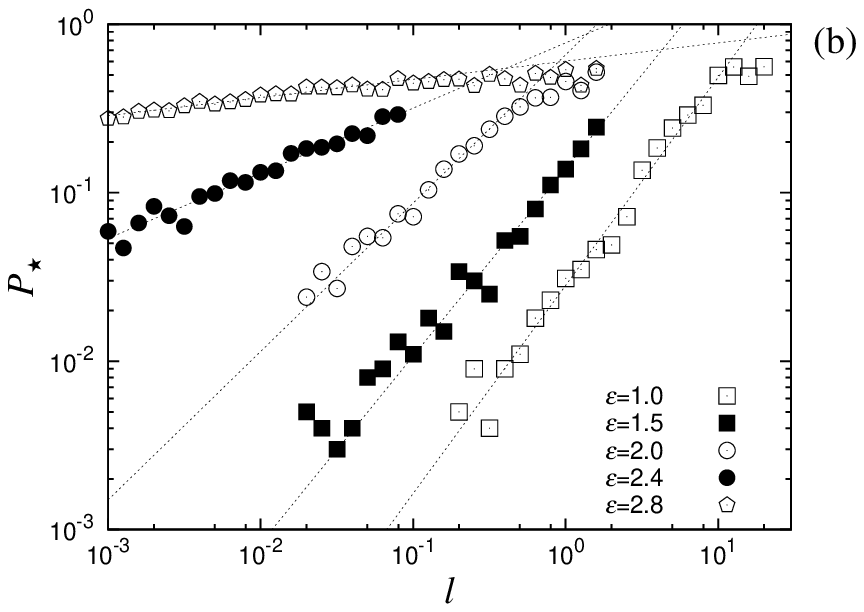}\\
\includegraphics[width=0.8\columnwidth]{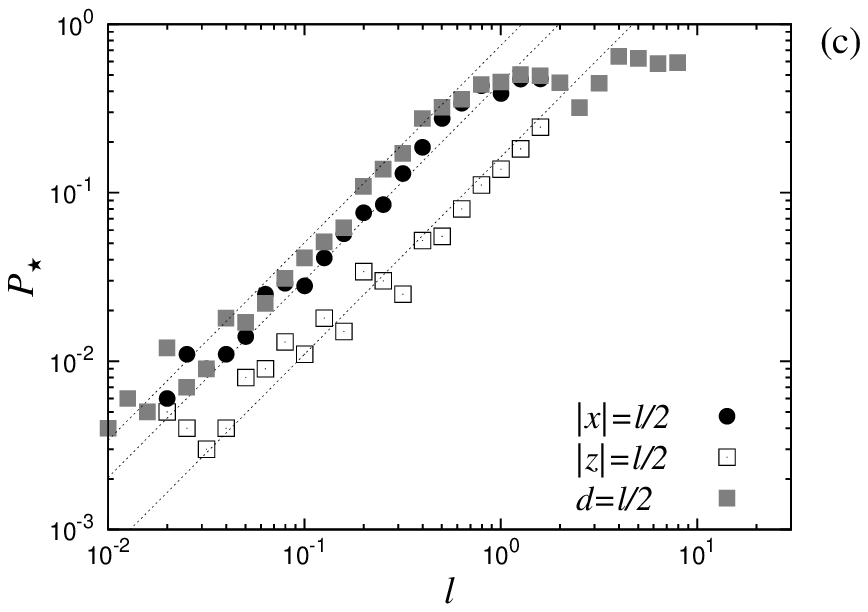}\\
\caption{\label{sca1} (a) Schematic figure showing the structure of
riddled basins near the invariant subspace that contains a chaotic
attractor. (b) Fraction of trajectories $P_\star$ that asymptote to the
antisynchronized state as a function of the distance $l=2|z|$ to the
synchronized state, for different values of the coupling strength $\varepsilon$. 
The full lines are least squares fits. 
(c) $P_\star$ for $\varepsilon=1.5$ and different orientations of the deviation $l$. } 
\end{figure}

This fine structure can be quantitatively characterized by the following
numerical experiment \cite{ott1,ott2}: let us consider the invariant manifold
at $x=y=z = 0$ and, depart from that manifold, for instance, 
by increasing $z$, up to a distance  $l=2|z|\equiv |z_1-z_2|$ 
[as depicted by the red line segment in  Fig. \ref{sca1}(a)]. 
Then we   evaluate the
fraction $V_l$ of points in that segment that belongs to the basin of ${\cal A}_s$. 
We obtained a numerical approximation of this fraction by considering a number of
initial conditions $x=y=0$, $|z|=l/2$, $X=Y=1.0$, and $Z$ randomly chosen in the
interval $[20,24)$. If the trajectories did not synchronize (within a small 
tolerance) up to a time  such that transients have elapsed and stationarity holds 
(typically, $t=10^3$), we consider that they asymptote to the antisynchronization attractor
${\cal A}_a$ and, accordingly, they  do not belong to the basin of ${\cal A}_s$. 
If the latter is riddled with tongues belonging to the basin
of ${\cal A}_a$, for any distance $l$ (no matter how small) there is always
a nonzero value of $V_l$. This fraction tends to zero as $l \rightarrow
0$. The fraction of length belonging to the basin of ${\cal A}_a$ 
(fraction of trajectories that do not synchronize) can be
written as $P_{\star} = 1 - V_l$, and is expected to scale with $l$ as a power
law $P_{\star}(l) \sim {l}^\eta$, where $\eta > 0$ is a scaling exponent. 
We integrated several initial conditions
at the same distance $l$ to the synchronization subspace and computed the
fraction $P_{\star}$ of initial conditions that do not synchronize, 
repeating this procedure varying the distance $l$. The results
shown in Fig.~\ref{sca1}(b) confirm the existence of a power law for this
fraction, for many values of the coupling strength. 

The results do not vary appreciably when  one departs from the 
synchronization manifold in other directions other than $z$.    
In Fig. \ref{sca1}(c) we plotted, for the same coupling strength ($\varepsilon=1.5$), 
the fraction $P_\star$ for initial conditions with $|z|=l/2$, $x=y=0$ (open squares),
and also for $|x|=l/2$, $y=z=0$ (filled circles) and  random values of $x,y,z$ such that 
$d\equiv\sqrt{x^2+y^2+z^2}=l/2$  (filled squares), obtaining essentially the 
same scaling exponent. 
This scaling behavior is observed within the interval $(\varepsilon_1,\varepsilon_2)$, 
below $\varepsilon_1$, no trajectories synchronize as seen in the previous sections, 
above $\varepsilon_2$, one observes a synchronized fraction but it does not change 
with $l$. 
The dependence of the numerically determined scaling exponents $\eta$ on the
coupling strength is depicted in Fig. \ref{compa}.
 
\begin{figure}[h]
\begin{center}
\includegraphics[width=0.9\columnwidth]{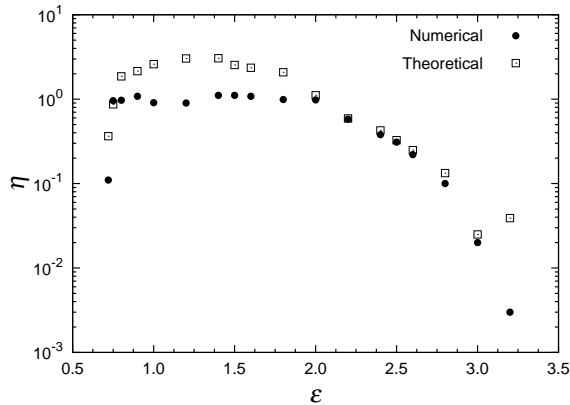} 
\end{center}
\caption{\label{compa} Scaling exponent $\eta$ for the fraction of
trajectories that asymptote to the antisynchronized state obtained by
a numerical experiment (filled circles). For comparison,  the   
theoretical values, given by $\eta=|\lambda_\perp|/D$  are also plotted (open
squares). 
} 
\end{figure}
An analytical expression for the exponent $\eta$ was derived by Ott and coworkers 
for a simple model (piecewise linear non-invertible map) \cite{ott2}. 
Their theoretical prediction arises from a diffusion approximation for a biased 
random walk that mimics the fluctuations of finite-time largest transversal 
Lyapunov exponents $\lambda_\perp(t)$. 
They obtain the law $P_\star \sim
{l}^\eta$, with $\eta = |\lambda_\perp|/D$ where $D$ is the diffusion coefficient.
This diffusion approximation is expected to be valid near the blowout 
bifurcation ($\lambda_\perp \simeq 0$) of an attractor with a riddled basin.  
The authors conjecture that a similar diffusion approximation, hence a similar 
relation involving parameters $\lambda_\perp$ and $D$,
 must rule the scaling relation in a large class of systems. 
The distributions shown in Fig. \ref{histo} already display a Gaussian character, 
which improves with larger time-$t$ interval,  
consistent with the probability distribution function of independent random 
innovations, that, by the central limit theorem,  is Gaussian. 
Additionally,   we plot in Fig.~\ref{dispersion}
the variance $\sigma^2_{\lambda_\perp(t)}$ of the probability distribution functions for
$\tilde{\lambda}_{\perp}(t)$ as a function of time, for different values of
the coupling strength. As a matter of fact,  the variance decays with time   
towards zero following asymptotically a power-law with exponent $-1$,  
as required for a normal diffusion processes, so validating 
the stochastic approach of Ott {\it et al.}. 
Accordingly, the diffusion coefficient $D$ can be estimated from the 
numerical curves, following $\sigma^2_{\lambda_\perp(t)} \sim 2D/t$.
The   estimates $\eta = |\lambda_\perp|/D$ are plotted in Fig. \ref{compa} together 
with the numerical values. Numerical and estimated values are in 
very good agreement in the proximity of the critical values, as expected \cite{andrade}. 
For intermediate values ($0.75<\varepsilon<2$) there is a discrepancy, 
and the numerical exponent remains close to one (linear behavior), as also 
observed for other systems with intermingled basins~\cite{tribolium}.

\begin{figure}[h] 
\includegraphics[width=0.9\columnwidth]{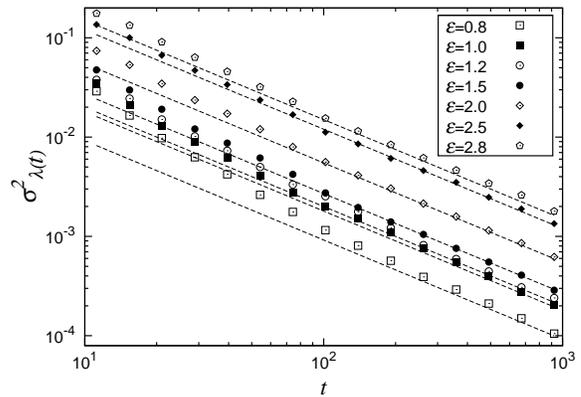}
\caption{\label{dispersion} Time decay of the variance of the
finite-time largest transversal Lyapunov exponent for different values
of the coupling strength $\varepsilon$. 
The lines are least squares fits 
of the function $f(t)=2D/t$ to the numerical points, for large $t$, 
allowing to estimate the diffusion coefficient $D$. 
}
\end{figure}

Another scaling law typical of riddled basins is related to the
 fraction of uncertain initial conditions, with respect to their final-state \cite{sabrina2}.  
We may regard riddled basins as an extreme case of fractal basins, for which
there is final-state  sensitivity and the uncertainty fraction  
scales as a power-law with the uncertainty level, whose exponent
gives a measure of the extreme final-state sensitivity due to riddling.
Consider again the points at $x=y=0$ and $|z| = l/2$  drawn in the phase space 
portrait in Fig. \ref{sca2}(a), as described earlier, and choose 
randomly an initial condition $\mathbf{x}_0$ on that region. 
Now choose randomly another initial condition $\mathbf{x}^\prime_0$ with
uniform probability within an interval of length  $2\xi$  and centered at
$\mathbf{x}_0$ [Fig. \ref{sca2}(a)]. If both points belong to different basins, they
can be referred to as $\xi$-uncertain \cite{ogy,ogy1}.

\begin{figure}[h] 
\begin{center}
\includegraphics[width=0.71\columnwidth]{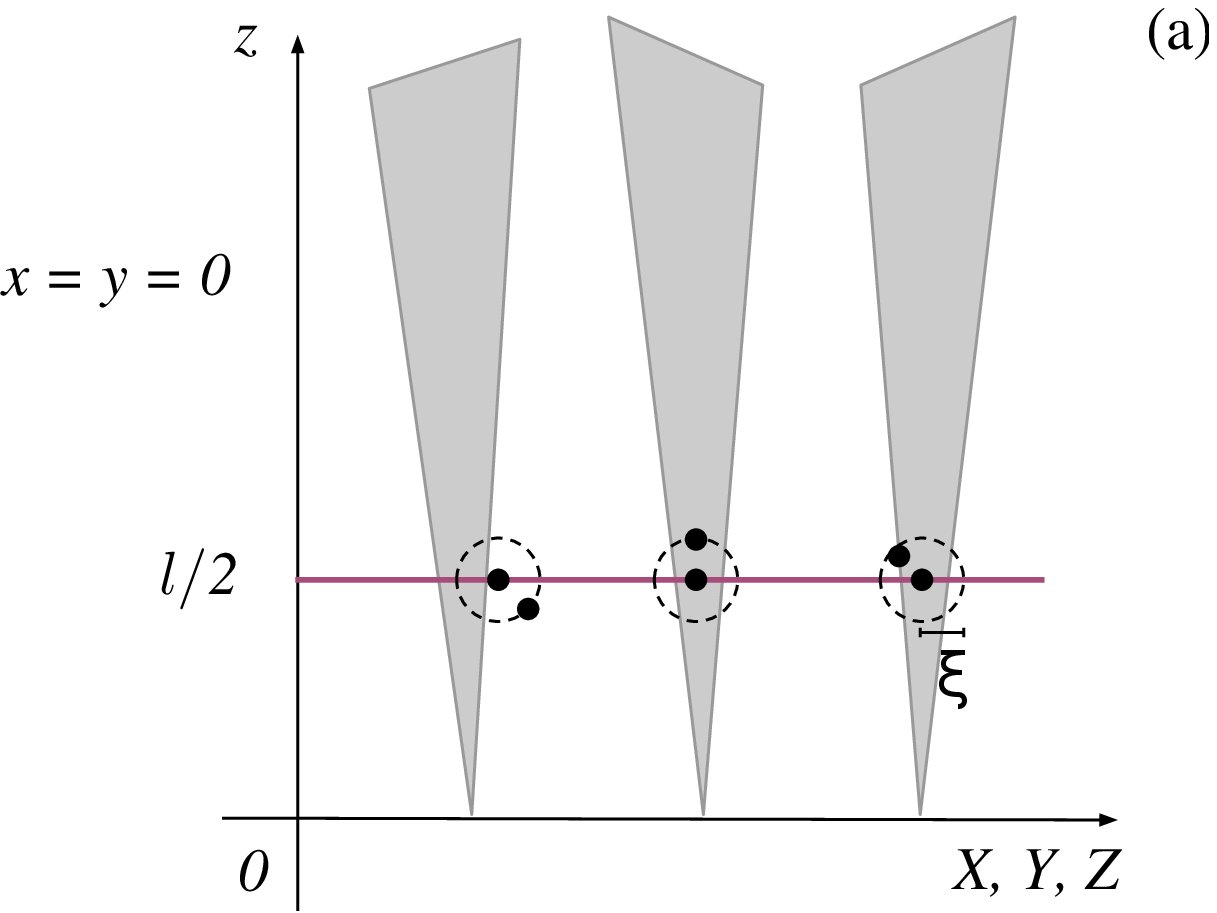} \\[5mm]
\includegraphics[width=0.75\columnwidth]{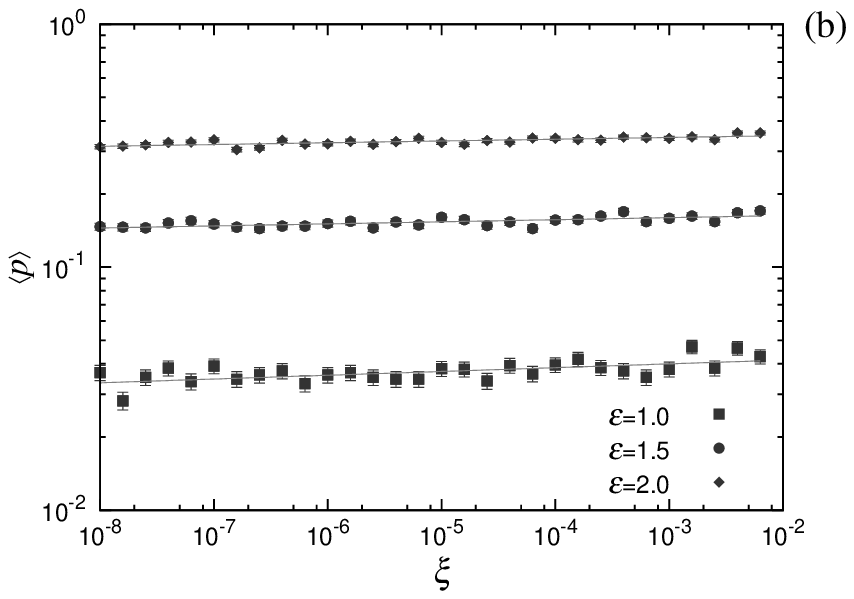} 
\end{center}
\caption{\label{sca2} (a) Schematic figure showing the numerical
determination of the uncertainty fraction. (b)
Fraction of uncertain initial conditions as a function of the
uncertainty level, for different values of the coupling strength. The
solid lines are least squares fits.} 
\end{figure}

The fraction of $\xi$-uncertain points, or uncertainty fraction,  
denoted by  $\langle p \rangle$, is the probability 
of making a mistake when attempting to
predict   which basin the initial condition  belongs to, given a
measurement uncertainty $\xi$. 
This probability scales with the
uncertainty level as a power law of the form $\langle p\rangle \sim
\xi^{\phi}$, where $\phi \ge 0$ depends on both $\mathbf{x}_0$ and $l$. 
Numerical results are shown in Fig. \ref{sca2}(b).
The stochastic model of Ott {\it et al.}  predicts a power-law, 
with exponent given by $\phi =\lambda_\perp^2/(4D\lambda_\parallel)$, 
a prediction that agrees with our numerical results close to the critical points. 
However, for intermediate values, while the stochastic model predicts 
small (though nonzero) values ($\phi<0.28$), 
our numerical results for $\phi$, as illustrated in Fig. \ref{sca2}(b),  
yield much smaller values (by a factor greater than ten).
As a matter of fact, for
riddled basins, the  exponent $\phi$ should be rigorously zero (i.e. there would be
no way to decrease the uncertainty fraction by decreasing the uncertainty
level). Indeed, our results [Fig. \ref{sca2}(b)] support this scaling law,
with numerically obtained exponents close to zero.

\section{Conclusions and final remarks}
\label{sec:final}

Riddled basins for the synchronization attractor of
coupled Lorenz systems have been previously suggested in the literature but
without a detailed characterization. In this work we offer numerical
evidence that, for a specified range  of the coupling parameter 
($\varepsilon_1 < \varepsilon < \varepsilon_2$), 
coupled Lorenz systems exhibit symmetrically riddled basins of attraction 
for synchronized and antisynchronized states. 
Since there are only two symmetric attractors, their basins are intermingled. 
We firstly  showed  that the
mathematical conditions for the existence of riddled basins are fulfilled, 
with the help of properties of finite-time largest transversal Lyapunov exponents and 
of the largest transversal exponent for particular orbits. This is important as 
furnishes the sources of local transversal instability of the attractor even if 
stable in average.
In a second place, we verified the existence of two scaling laws
characterizing quantitatively the degree of uncertainty related to the
riddled basins.  These numerical results were compared to an analytical 
prediction (the stochastic model   \cite{ott2}), yielding a good accord where 
expected.   Beyond the characterization of the structure of a riddled basin, 
these scaling laws allow to quantify the limitations to improve the ability in determining 
the final state of the system by increasing the accuracy level.

Let us remark that intermingling, in particular of symmetric basins, 
has  also been observed in 
other systems with either continuous (e.g., mechanical system \cite{on-off} and 
coupled R\"ossler oscillators\cite{yanchuk01})  or discrete time dynamics 
(coupled logistic maps \cite{maistrenko}).  
In the latter case, the analysis of the lowest period orbit was enough to 
furnish the conditions for the occurrence of riddling in certain parameter region. 
In fact,   as anticipated by the results presented in Fig. \ref{fig:upos}, 
deepening in that point of view 
may  furnish precise information on the nature of the bifurcations  
triggering riddling, although this may be a difficult task for the present system. 
As other perspectives for future work on this system, let us also mention the 
plausible occurrence, beyond the blowout bifurcation,  
of  two-state on-off  intermittency \cite{on-off} 
 for which there is some evidence  \cite{kim}. 
Finally, it can still be worthy to explore  other regions of phase space, 
as well as other ranges (negative or large values) of the coupling parameter.

In any case, for applications, multistability 
is already a source of troubles.  
Still worst, the existence of intermingled basins of attraction for the synchronized and
antisynchronized chaotic states of this system jeopardizes the solution of
the problem of ensuring a given final state, since the initial condition 
determination is always done within
a certain uncertainty level. With riddled basins, any uncertainty level,
however small, lead to complete indeterminacy of the future state of the
system. Hence in this case we cannot use synchronization of chaos for any
practical purpose, since we will always be haunted by the existence of the
another, antisynchronized state, with a basin intermingled with the basin of
the synchronized state. 
Of course, the same difficulties concern  the predictability of natural 
phenomena modeled by coupled Lorenz systems. Therefore, the importance of 
detecting the regimes where riddling can occur in a dynamical system.

\section*{Acknowledgements:}

This work was made possible with help of CNPq, CAPES, FAPERJ, and Funda\c{c}\~ao
Arauc\'aria (Brazilian Government Agencies).


\begin{thebibliography}{99}
\bibitem{exigence} Notice that, due to the own nature of the basins, 
there are hindrances to determine them numerically. 
We considered that an orbit asymptotes to the attractor ${\cal A}_s$ 
(${\cal A}_a$) if $|x|,|y|,|z|<10^{-8}$ 
($|X|,|Y|,|z|<10^{-8}$) 
and verified that the resulting picture is not altered for later times.
\end{thebibliography}
\end{document}